\input{aipcheck}

\documentclass[
    ,final            
  ]
  {aipproc}

\layoutstyle{6x9}

\begin{document}

\title{Neutron Star Astronomy in the era of the European Extremely Large Telescope}


\author{Roberto P. Mignani}{address={Mullard Space Science Laboratory - University College London}}

\begin{abstract}
About 25 isolated neutron stars (INSs)  are now detected in the optical domain, mainly  thanks to the {\em HST} and  to {\em VLT}-class telescopes.  The {\em European Extremely Large Telescope} ({\em E-ELT}) will yield $\sim 100$ new  identifications, many of which  from the follow-up of {\em SKA}, {\em IXO}, and {\em Fermi} observations.  Moreover, the {\em E-ELT} will allow to carry out, on a much larger sample, INS observations which still challenge  {\em VLT}-class telescopes, enabling studies on the structure and composition of the NS interior, of its atmosphere and magnetosphere, as well as to search for debris discs. In this contribution, I  outline future perspectives for NS optical astronomy with the {\em E-ELT}. 
\end{abstract}

\maketitle

\section{Introduction}

While INS studies in the optical domain started with the identification of the Crab pulsar in 1969,  followed with a $\sim 10$ year period by the identification of the Vela pulsar and the discovery of optical pulsations from PSR\, B0540$-$69, it was only at the beginning of the 1990s that  they deserved  a dedicated niche, thanks both to the deployment of the ESO {\em NTT} 
and the launch of the {\em HST}.
As a result, the first optical identification of the fainter middle-aged/old pulsars were achieved, coupled with the first detections of INSs in the near-UV and near-IR, and with the first optical identifications of the X-ray Dim INSs (XDINSs). Moreover, the first INS detections in the extreme-UV were obtained by the {\em EUVE}.   A decade later, the advent of the 8m-class telescopes, like the {\em VLT} (Mignani 2009a), allowed to build on these results. While {\em HST} observations made optical astrometry a default tool to obtain  proper motion and parallax measurements  for both radio-silent INSs and radio pulsars, observations with 8m-class telescopes allowed to obtain the first spectroscopy and polarimetry measurements for INSs fainter than $m_{V} \sim 22$,  to discover the magnetar IR counterparts and to measure their optical/IR pulsations.   {\em Spitzer} complemented on the work of the {\'em EUVE} by allowing to detect mid-IR emission from an INS other than the Crab and to constrain the emission properties for several of them.  So far, a total of possibly 24 INSs have been identified at UV/optical/IR (UVOIR) wavelengths, including rotation-powered pulsars, XDINSs, and magnetars (e.g., Mignani 2009b, 2010a). As number matters,  the UVOIR domain  occupies now more than a niche  in the panorama  of multi-wavelength observations of INSs.  
Being intrinsically the faintest stars, INS optical studies obviously require a huge collecting power. While 8m-class telescopes yielded an $\sim 100\%$  increase in detection rate with respect to mid-size telescopes like the 3.5m {\em NTT}, they allowed to perform deep studies (timing, spectroscopy, and polarimetry) only for a tiny fraction of the optically identified INSs.
Indeed, although medium-resolution spectra have been obtained for 7 INSs,  only for the Crab they have an appreciably high signal--to--noise.  Thus, in most cases the study of the UVOIR spectral flux distribution relied on multi-band photometry measurements, often taken on time scales of years. Measurements of the linear polarisation has been obtained only for 3 pulsars but, once again, the Crab is the only one for which repeated measurements exist (both phase-averaged and phase-resolved). On the other hand, pulsations have measured for only one third of the sample (Mignani 2010b), most of them thanks to the {\em HST} and the {\em VLT}.  

\section{The European Extremely Large Telescope}

New observing facilities are thus required to improve, both in quantity and quality, the existing sample. 
In the next decade many of such observing facilities will be deployed. The {\em HST} will be ideally replaced in 2015 by the {\em James Webb Space Telescope (JWST)},
with a mirror of 6.5m the largest optical space telescope ever launched. On the ground,  a new generation of giant telescopes will follow, e.g. the {\em Giant Magellan Telescope (GMT)} 7$\times$8.4m mirror array,
the   {\em Thirty Meter Telescope (TMT)},
and the 42m  ESO {\em E-ELT}.
With a collecting area of 1200 m$^2$, compared with the 50 m$^2$ of the {\em VLT}, the {\em E-ELT} will outsize by a factor of 2 and 3 its ground-based competitors. In particular, the {\em E-ELT} will provide a maximum field--of--view of 10', a 0.3-24 $\mu$m wavelength coverage, and, thanks to the adaptive optic correction, a spatial resolution of 5 mas at 1 $\mu$m, better than the 34 mas resolution of the {\em JWST} at the same wavelength. The {\em E-ELT}, to be built on the top of Cerro Armazones in Chile,  will be equipped with a suite of 9 instruments, whose development is currently in Phase A.
Although the {\em E-ELT} instrument suite has been designed primarily for  extra-galactic and stellar astronomy and it is optimised for longer wavelengths,  a few instruments can be successfully employed for NS astronomy.  In the optical, the most useful instrument is  {\em OPTIMOS}, an imager/spectrograph (0.37-1.6 $\mu$m) with a large field--of--view of $6.78' \times 6.78'$  (50 mas pixel size), while {\em EPICS} (0.6-0.9 $\mu$m) can be used for imaging polarimeter of point sources over a $2"\times2"$ field--of--view. In the near-IR, the reference instrument is {\em MICADO}, an imager/spectrograph  (0.8-2.5 $\mu$m) with a spatial resolution as sharp as 3 mas per pixel over a field--of--view of $53" \times53"$. In the mid-IR, {\em MICADO} is complemented by {\em METIS}, also an imager/spectrograph  (2.9-14 $\mu$m), with a smaller field--of--view ($17.6"\times17.6"$) and a relatively lower spatial resolution (20 mas per pixel).  While these instruments are suited for imaging, spectroscopy, and polarimetry observations, none of them is expressly designed for timing observations.
Being INSs  multi-wavelength pulsars, this might represent a severe  limitation of the {\em E-ELT}  potentials for NS astronomy. As in the case of  other instruments at the {\em VLT}, however, timing measurements can be obtained from imaging observations through fast detector readout modes. Of course, this technique can be applied  for the brightest  targets (not much of a problem for the {\em E-ELT}), and the period windows which can be explored depends on the  readout times. Alternatively, timing detectors can be plugged-in to planned instruments, for instance fed from the fibre systems of {\em MICADO} and {\em OPTIMOS}.   While dedicated timing instruments might still be developed before Phase B,  guest instruments  can offer an important alternative in wait for second generation {\em E-ELT} instruments. An interesting concept design is that of  {\em QuantEYE} (e.g., Barbieri et al. 2010), a detector with a nsec photon time-tag accuracy, whose prototype ({\em IquEYE}) has been already tested at the {\em NTT}.  Interestingly enough, timing observations with the {\em E-ELT} gain momentum in the astronomical Community (Shearer et al. 2010), with the creation of a dedicated scientific forum ({\it http://faststar.freeforums.org/}).


\section{Scientific impact}

The {\em E-ELT} will have the potentials of detecting INSs which are fainter, further away, and more absorbed and to carry out, on a much shorter time span, deeper studies,  only explored with 8m-class telescopes. These studies will help to address many open issues in NS astronomy (Mignani 2010c), to study the structure an composition of the NS interior, to investigate the presence of an atmosphere, to study the properties of the magnetosphere, to search for debris discs, and to understand the NS formation an evolution, hence the NS variety.  
First of all, its deep imaging capabilities (down to $m_V \sim 32$), wide spectral range, and sharp  spatial resolution will allow to search for INS emission without being biased by the source spectrum, and by the field confusion in crowded regions like, e.g. in the galactic plane where most magnetars have been discovered.   Radio-loud INSs still represent the majority of the population, with up to $\sim$ 2500 and $\sim 30000$ new pulsars to be discovered with on-going radio survey and with the {\em SKA}.  Thus, they represent a primary target for {\em E-ELT} imaging observations, since radio observations yield sub-arcsec positions, DM distance estimates, and energetics from which expectation values for the optical flux can be derived. More targets, especially radio-silent INSs, will come from high-energy observations performed with {\em IXO} and {\em Fermi}, with optical observations crucial to understand the nature of these sources. About 100 INSs are, thus, expected to be identified with the {\em E-ELT}. Moreover, optical/IR imaging with the {\em E-ELT} will be the only way to carry out INS astrometry measurements, with {\em MICADO} able to measure proper motions up to the LMC and parallaxes at distances above 1 kpc. The detection of the faint  optical bow-shocks formed by the INS  motion in the ISM will then enable to infer the INS radial velocity and  determine its space velocity which is crucial to reconstruct its galactic  orbit, to localise its birth place and identify the progenitor star clusters. 
Optical spectroscopy with {\em OPTIMOS} will be crucial to map the NS surface temperature, unveiling colder areas not detectable in X-rays and produced by an uneven heat transfer in the NS interior, and to measure the bulk surface temperature of $> 10$ Myear old and colder NSs, which is critical to constrain the tails of model cooling curves. Optical spectroscopy  will also allow to search for distortions in the radiation spectrum from the NS surface induced by the presence of an atmosphere, including absorption features from which one can derive the composition of the atmosphere and its ionisation level, density, opacity, gravitational red-shift. Optical and IR spectroscopy with {\em OPTIMOS},  {\em METIS}, and {\em MICADO} will allow to study the emission spectrum from the NS magnetosphere, to search for possible cyclotron features, and discover multiple spectral breaks which can be related to complex particle distributions. Moreover, IR imaging/spectroscopy  with {\em METIS} will allow to detect debris discs, to constrain their mass, size, and temperature, and investigate possible interactions with the INS (torque, accretion). Discovering debris discs will be important to test supernova explosion models, to trace the early post-supernova phases, and  investigate planet formation.  Last but not least, optical/IR {\em E-ELT} spectroscopy will allow to study the properties of stars in the NS parental clusters up to distances of  $>5$ kpc and $A_V>10$ and to determine whether different INS characteristics depends on the progenitor mass, as suggested for the magnetars. 
Timing observations would allow to identify many more pulsars up to the M31 distance,  study their optical light curves with an unprecedented level of details, study the wavelength dependence of the light curve, and test magnetosphere emission models. Moreover, timing with the {\em E-ELT} is the only way to carry out systematic searches for giant optical pulses and to study their link with the giant radio pulses, hence the connection between incoherent and coherent radiation. IR timing would also allow to verify the presence of debris discs by isolating pulsed and DC emission components, presumably produced from the NS and from the disc, and to find evidence of disc reprocessing of the NS X-ray pulse from observed phase lags and pulse fraction difference.  
Polarimetry with the {\em ELT}, either with {\em EPICS} or with a more general-purpose instrument, would help to determine the polarisation degree of the NS atmosphere, to test models of the NS magnetosphere, to constrain the magnetic field angle with respect to the NS spin axis and its orientation  in the sky, to resolve the polarisation map in PWNe  and SNRs, to measure magnetar burst collimation from the polarisation position angle, and study the magnetic field evolution after a burst.
More possibilities for NS studies  might hopefully emerge from second generation {\em E-ELT}  instruments by combining different  observing techniques. For instance, phase-resolved spectroscopy 
would allow to better map  the NS surface temperature as the star rotates, as well as to map the particle distribution in different regions of the NS magnetosphere. Phase-resolved polarimetry would allow to map the NS magnetic field geometry following the variation of the polarisation degrees  along the pulsar period.  Spectro-polarimetry would allow to measure the magnetic fields of magnetar parental cluster stars from the Zeeman line splitting and to determine whether they, instead,  form from the collapse of hyper-magnetic ($>$ 1 kG) progenitors.

After the pioneering years, optical obsevations are now a main field in INS studies, being crucial to complete their multi-wavelength phenomenology, address many issues on NS physics, understand the INS diversity, and draw a Grand Unification scheme of their formation and evolution. This requires in-depth investigations, just explored so far. Only the future ELTs can provide this opportunity, opening a new era in NS astronomy.


%
 



\begin{theacknowledgments}
RPM acknowledges  OPTICON for supporting his participation to ASTRONS 2010  
\end{theacknowledgments}



\bibliographystyle{aipproc}   

\bibliography{sample}

\IfFileExists{\jobname.bbl}{}
 {\typeout{}
  \typeout{******************************************}
  \typeout{** Please run "bibtex \jobname" to optain}
  \typeout{** the bibliography and then re-run LaTeX}
  \typeout{** twice to fix the references!}
  \typeout{******************************************}
  \typeout{}
 }


\end{document}